\documentclass[12pt]{iopart}
\usepackage{iopams,psfig,amssymb}
\eqnobysec

\begin{document}
\jl{1}

\title[High-temperature expansion for quasiperiodic Ising models]
{High-temperature expansion for Ising models on quasiperiodic tilings}

\author{Przemys{\l}aw Repetowicz, Uwe Grimm and Michael Schreiber}

\address{Institut f\"ur Physik,
         Technische Universit\"at, 
         D-09107 Chemnitz, 
         Germany}

\begin{abstract}
We consider high-temperature expansions for the free energy of
zero-field Ising models on planar quasiperiodic graphs.  For the
Penrose and the octagonal Ammann-Beenker tiling, we compute the
expansion coefficients up to $18$th order. As a by-product, we obtain
exact vertex-averaged numbers of self-avoiding polygons on these
quasiperiodic graphs. In addition, we analyze periodic approximants by
computing the partition function via the Kac-Ward determinant. For the
critical properties, we find complete agreement with the commonly
accepted conjecture that the models under consideration belong to the
same universality class as those on periodic two-dimensional lattices.
\end{abstract}


\submitted

\maketitle

\section{Introduction}
\label{sec1}

Since the discovery of quasicrystals in the early eighties
\cite{Schechtman,IshNisFuk,Bendersky,WanCheKuo} considerable attention
has been paid to the magnetic properties of these materials. While
many quasicrystals contain atoms (such as Fe, Mn, or rare-earth
elements) that carry local magnetic moments, these are usually
screened very effectively, and consequently one finds a weak
paramagnetic or diamagnetic behaviour, see e.g.~\cite{Berger,Ath}.
Recently, however, there has been ample experimental evidence for
magnetic ordering in quasicrystals, including ferrimagnetic
\cite{YokInoMas}, ferromagnetic \cite{LyuLinLin}, anti-ferromagnetic
\cite{ChaOulSch}, and spin-glass behaviour
\cite{GAVCO,NKWSWSF,PSSIYFS,SimHipAudTdL}, though some results are
still discussed controversially, see e.g.~\cite{IFZCSG,STTS}, in
particular with regard to the importance of crystalline phases present
in the samples.

Even before magnetic ordering in quasicrystals had been observed
experimentally, theoretical investigations on the influence of
quasiperiodic order on magnetic properties commenced. In most cases,
the models considered were either one-dimensional quantum spin chains
with aperiodic sequences of coupling constants or classical Ising
models on two-dimensional quasiperiodic graphs; we refer the reader to
the recent review \cite{GriBaa} for a rather complete list of
references. Recently, a symmetry classification scheme for
magnetically ordered quasicrystals has been proposed \cite{Ron}.

In this context, it is one of the central questions whether
quasiperiodic order influences the universal properties at the phase
transition, such as the critical exponents, in comparison to the
periodic case.  There is a heuristic criterion due to Luck \cite{Luck}
on the relevance of aperiodicity.  According to this criterion, the
``topological disorder'' encountered in two-dimensional quasicrystals,
generated by the cut-and-project method, is irrelevant; and hence an
Ising model on a quasiperiodic tiling should belong to the same
universality class as the Ising model on the square lattice. Clearly,
non-universal properties do in general depend on the particular system
under consideration. For instance, the location of critical points of
lattice models depends in a systematic way on the structure of the
graph on which the model is defined, see \cite{BaaGriRepJos} and
references therein.

In this article, we consider high-temperature expansions of the free
energy for zero-field Ising models on two planar quasiperiodic graphs,
the decagonal Penrose \cite{Penrose,deBruijn} and the octagonal
Ammann-Beenker \cite{AmmGruShe,DunMosOgu,Katz} tiling. The technique
of high-temperature expansions is well known, see e.g.~\cite{Domb}; it
was developed several decades ago and has since been applied to a
variety of periodic lattices in both two and three dimensions. With
regard to previous work on high-temperature expansions of
quasiperiodic Ising models, we are only aware of two articles by Abe
and Dotera \cite{Abe,Dotera} who compute the expansion of the free
energy up to the eighth order for the Penrose tiling and its dual, and
of a few numerically calculated expansion coefficients for the
susceptibility for the Penrose case \cite{Miyazima}.  Employing a
systematic procedure, we are able to compute the exact values of the
coefficients up to the 18th order for both the Penrose and the
Ammann-Beenker tiling. This requires much more effort than the
calculation for periodic lattices, because the number of graphs that
one has to take into account grows tremendously with the
order. Although our expansions are still not yet sufficient to extract
good estimates for the critical temperatures or the critical
exponents, we can show that our results are consistent with those
obtained by different methods.

Presently, the most accurate data on the transition temperature and
the critical exponents stem from Monte-Carlo simulations
\cite{OkaNii1,OkaNii2,Sorensen,Ledue}.  Besides graphical expansions
and Monte-Carlo simulations, further methods have been employed to
gain information about the critical behaviour of quasiperiodic Ising
models. First of all, exactly solvable cases can be constructed as,
for instance, the Ising model on the so-called labyrinth tiling
\cite{BaaGriBax}, see also \cite{GriBaa,Choy} for further
examples. These models correspond to particular choices of coupling
constants, restricted by the requirement of integrability, and thus
might not be representative for the general situation.  For the
solvable models based on the idea of ``$Z$-invariance'', see
\cite{GriBaa} and references therein, the critical behaviour
necessarily is the same as for the periodic case, but one does not get
a clue whether this extends to the general case, or whether it is at
least the generic situation.  Secondly, there is an interesting
approach using Lee-Yang zeros \cite{Lee}, which are complex roots of
the partition function in certain variables.  Simon and Baake
\cite{SimBaa,Simon} calculated the zeros of the partition function for
a large patch of the Ammann-Beenker tiling numerically and drew
conclusions about the critical temperature and the critical exponents.
Furthermore, renormalization group techniques were applied to study
the Ising model on two-dimensional quasiperiodic tilings
\cite{AoyOda}. In that case, one exploits the self-similarity of
quasiperiodic tilings, which translates into a renormalization
procedure that, however, can only be treated approximately in general.
We note that for one-dimensional quantum Ising chains with
aperiodically modulated coupling constants, corresponding to
two-dimensional layered Ising models, renormalization techniques may
yield exact results for the critical behaviour
\cite{HerGriBaa,HerGri}. In this case, the modulation is
one-dimensional, and in accordance with Luck's criterion \cite{Luck}
one finds that the critical behaviour depends on the fluctuations of
the aperiodic sequence of coupling constants.

So far, all results appear to be in accordance with Luck's criterion,
including a recent Monte-Carlo study of the three-state Potts model on
the Ammann-Beenker tiling \cite{LBLT}. Still, most approaches are
based on numerical or approximative treatments. In contrast, it is our
aim to obtain exact values of the coefficients for the
high-temperature expansions in the present paper, which is organized
as follows. In the subsequent section, we briefly recall the graphical
high-temperature expansion of Ising models. In \sref{sec3}, we discuss
the generation of subgraphs of quasiperiodic tilings and the
computation of their occurrence frequencies. Then, in \sref{sec4}, we
present our results for the coefficients of the high-temperature
expansions for the Penrose and the Ammann-Beenker tiling. The
corresponding implications for the critical behaviour are discussed in
\sref{sec5}. In \sref{sec6}, we compare our results with exact
calculations of the partition functions of periodic approximants.
Finally, in \sref{sec7}, we present our conclusions.

\section{High-temperature expansion}
\label{sec2}

We now give a brief account of the high-temperature expansion for the
free energy of an Ising model on a graph without an external field
\cite{Domb}.  Let us consider a finite graph ${\cal G}$ containing $N$
sites (vertices) with $M$ neighbour pairs of vertices connected by
bonds.  We emphasize that, throughout the paper, the notion of
neighbouring vertices refers to vertices connected by a bond, and not
to the geometric distance between the vertices. For instance, in the
Penrose tiling discussed below, the short diagonal of the small rhomb
corresponds to the smallest distance between vertices, but does not
constitute a bond. At a vertex $j$, we place an Ising spin
$\sigma_{j}\in\{\pm 1\}$; and two spins $\sigma_{j}$ and $\sigma_{k}$
located at neighbouring vertices $j$ and $k$ interact with a coupling
constant $J$ which we assume to be independent of the position.  Hence
the energy of a spin configuration
$\bsigma=\{\sigma_1,\sigma_2,\ldots,\sigma_N\}$ on ${\cal G}$ is given
by
\begin{equation}
E(\bsigma) \; =\;  
-J\sum_{\langle j,k\rangle} \sigma_{j}\sigma_{k}
\end{equation}
where we sum over all neighbour pairs $\langle j,k\rangle$ connected
by bonds as mentioned above.  The logarithm of the partition function
\begin{eqnarray}
Z({\cal G}) &  = &  \sum_{\bsigma}
\exp\left[-\beta E(\bsigma)\right] \nonumber\\ 
& = & \left[\cosh (\beta J)\right]^{M} \sum_{\bsigma}
\prod_{\langle j,k\rangle}
\left[ 1+\sigma_{j}\sigma_{k}\tanh (\beta J)\right] 
\end{eqnarray}
is, apart from a factor $-1/\beta$, the free energy. It can be
expanded as
\begin{equation} 
\ln{Z({\cal G})} \; =\; N\ln{2} +
M\ln{[\cosh{(\beta J)}]} + 
N\sum\limits_{n=1}^{\infty} g_n\, w^n \; ,
\label{eq:expansion1}
\end{equation} 
where $\beta = 1/k_{B} T$ with Boltzmann's constant $k_{B}$ and
temperature $T$. The expansion variable
\begin{equation}
w \;=\; \tanh (\beta J) \; ,
\end{equation} 
is small for high temperature, hence the notion high-temperature
expansion.  The expansion coefficients $g_n$ are related to the number
of subgraphs of ${\cal G}$ containing $n$ bonds.

The terms in the expansion \eref{eq:expansion1} can be rearranged in a
different fashion which is more convenient for our needs (see page 382
in \cite{Domb})
\begin{eqnarray}
\ln{\tilde{Z}({\cal G})} & = &
\ln{Z({\cal G})} - N\ln{2} - M\ln{[\cosh{(\beta J)}]} \nonumber\\
& = & N\sum\limits_{n=1}^{\infty} g_n\, w^n 
\; = \; \sum_{r} (c_r ;{\cal G})\, k_r(w) \; ,
\label{eq:expansion2}
\end{eqnarray}
where we now sum over all {\em connected}\/ subgraphs $c_r$ of ${\cal
G}$.  The quantity $(c_r ;{\cal G})$ denotes the so-called {\em
lattice constant}\/ of $c_r$ in ${\cal G}$, counting the number of
ways $c_r$ can be embedded in ${\cal G}$.  The weight functions
$k_r(w)$ depend only on $c_r$, not on ${\cal G}$. In our case, without
external field, we can restrict the sum to so-called {\em star
graphs}\/. These are graphs that cannot be dissected into two disjoint
subgraphs by eliminating a single vertex. 

The weight functions $k_r(w)$ in equation \eref{eq:expansion2} can be
calculated from the partition function $\tilde{Z}(c_r)$ of the
subgraph $c_r$. For this aim, let us generate all star subgraphs and
arrange them in a sequence $\{c_{r}\}_{r=1,2,\ldots}$ such that $c_s$
cannot be embedded in $c_{r}$ for $r<s$. In other words, the lattice
constant $(c_s;c_r)$ may be non-zero only if $s\le r$, which, in
general, does not determine the sequence uniquely. Having arranged the
subgraphs in such a way, the expansion \eref{eq:expansion2} for a
subgraph $c_r$ gives
\begin{equation}
\ln{\tilde{Z}(c_r)} \;=\; \sum_{s=1}^r (c_s;c_r)\, k_s(w) 
\end{equation}
and, taking into account that $(c_r;c_r)=1$, we obtain the
corresponding weight $k_r(w)$
\begin{equation}
k_r(w) \;=\; \ln{\tilde{Z}(c_r)} \;-\; \sum_{s=1}^{r-1} (c_s;c_r)\, k_s(w) 
\label{eq:weights}
\end{equation}
expressed in terms of lattice constants $(c_s;c_r)$ and weights
$k_s(w)$ with $s<r$. Therefore, we can compute the weights $k_r(w)$
successively provided we know the partition function $\tilde{Z}(c_r)$
and the lattice constants $(c_s;c_r)$ of all star graphs $c_s$ that
are subgraphs of $c_r$. 

We note that we can rearrange the sum in equation \eref{eq:expansion2}
as
\begin{equation}
\ln{\tilde{Z}({\cal G})}\;
 = \;\sum_{n=3}^{\infty}\sum_{r}\sum_{s}\,
(c^{(n)}_{r,s};{\cal G})\, k^{(n)}_{r,s}(w)
\label{eq:expansion2l}
\end{equation}
where $r$ labels closed loops $l^{(n)}_r$ consisting of $n$ bonds, and
$c^{(n)}_{r,s}$ are all possible complete ``fillings'' of the loop
$l^{(n)}_r$.  By ``fillings'' of a loop we mean all proper subgraphs
of ${\cal G}$ which have the loop as their boundary. Here, the
functions $k^{(n)}_{r,s}(w)$ have the form
\begin{equation}
k^{(n)}_{r,s}(w) \; = \; w^n + \Or (w^{n+1}) \; ,
\end{equation}
hence truncating the sum over $n$ in equation \eref{eq:expansion2l}
yields all terms in the expansion up to $n$th order in $w$. The
calculation of the weight functions $k_{r,s}^{(n)}(w)$ can be
performed in analogy to that of the weight functions $k_{r}(w)$
\eref{eq:weights}.

In summary, in order to calculate the high-temperature expansion
\eref{eq:expansion2l} of the Ising model to order $n_{\rm max}$ we have to 
perform the following steps:
\begin{enumerate}
\item generate all loops $l^{(n)}_r$ in the graph ${\cal G}$ consisting of
      $n\le n_{\rm max}$ bonds;
\item construct all fillings $c^{(n)}_{r,s}$ of $l^{(n)}_r$;
\item calculate $\ln{\tilde{Z}(c^{(n)}_{r,s})}$, the logarithm of the 
      partition function for the subgraphs $c^{(n)}_{r,s}$;
\item calculate the lattice constants $(c^{(n)}_{r,s};{\cal G})$ 
      and $(c^{(n')}_{r',s'};c^{(n)}_{r,s})$;
\item compute the weight functions $k^{(n)}_{r,s}(w)$ by successive use of
      the analogue of \eref{eq:weights};
\item calculate the expansion \eref{eq:expansion2l}.
\end{enumerate}
We are now in the position to apply this scheme to the case of
quasiperiodic graphs. 

\section{Frequencies of subgraphs of quasiperiodic tilings}
\label{sec3}

In fact, we want to obtain the expansion \eref{eq:expansion2} for the
Ising model on an infinite quasiperiodic graph ${\cal G}$. Therefore,
we have to compute the corresponding ``averaged lattice constants'' per
vertex
\begin{equation} 
\langle c_r;{\cal G}\rangle := 
\lim_{N\rightarrow\infty} \frac{1}{N} (c_r;{\cal G}_N)
\end{equation}
where ${\cal G}_N$ denotes finite patches with $N$ vertices
approaching the infinite graph ${\cal G}$. In other words, we need to
calculate the occurrence frequency of a subgraph $c_r$ in the infinite
graph ${\cal G}$. The main challenge now is to compute these
quantities for a given quasiperiodic graph and all of its subgraphs up
to a certain size.

\begin{figure}
\centerline{\psfig{figure=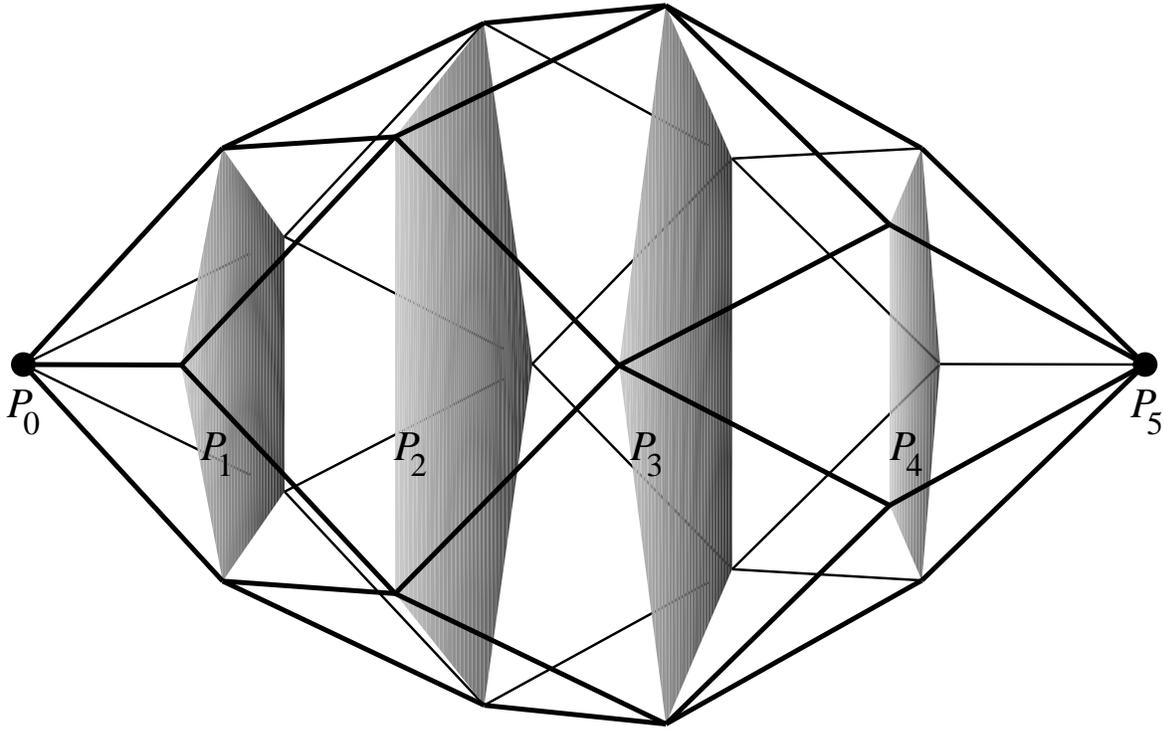,width=\textwidth,angle=-90}}
\caption{The acceptance domain of the Penrose tiling, consisting of
         four regular pentagons $P_{1}$, $P_{2}$, $P_{3}$, $P_{4}$, 
         and two isolated points $P_{0}$, $P_{5}$, situated on 
         equidistant parallel planes in the three-dimensional space 
         $E_{\perp}$. The polytope spanned by the lines is the projection
         of the five-dimensional hypercube to $E_{\perp}$.
         \label{fig:pentagons}}
\end{figure}

For quasiperiodic graphs generated by the cut-and-project method
\cite{DunKatz} the frequencies of subgraphs can be computed
exactly. In the cut-and-project method, one starts from a
higher-dimensional periodic lattice, and projects a certain part of it
onto a lower-dimensional ``physical'' or ``parallel'' space
$E_\parallel$.  For the two cases of interest, the Penrose and the
octagonal Ammann-Beenker tiling, the lattices have to be at least
four-dimensional, the minimal choice being the root lattice $A_4$ for
the Penrose case \cite{BKSZ} and the hypercubic lattice $\mathbb{Z}^4$
for the octagonal case \cite{DunMosOgu}. The root lattice $A_4$ can be
considered as a sublattice of $\mathbb{Z}^5$, wherefore the latter,
albeit not minimal, is frequently used to generate the Penrose
tiling. The physical space $E_{\parallel}$ is determined as an
invariant subspace with respect to the relevant subgroup (in our
examples the dihedral groups $D_5$ and $D_8$, respectively) of the
point group of the periodic lattice. Its orthogonal complement, the
perpendicular space $E_{\perp}$, is then also an invariant subspace of
this symmetry.  The quasiperiodic tiling is now obtained by projecting
all those lattice points onto $E_{\parallel}$ whose projection onto
$E_{\perp}$ falls into a certain set called the ``window'' or
``acceptance domain'' $A$. In the minimal case, this acceptance domain
has the same dimension as $E_{\perp}$; however, if we project the
Penrose tiling from the hypercubic lattice $\mathbb{Z}^5$, the
perpendicular space is three-dimensional and the acceptance domain
consists of four regular pentagons $P_m$ ($m=1,2,3,4$) situated on
equidistant, parallel planes, and two isolated points ($P_0$ and
$P_5$), see \fref{fig:pentagons}. For the Ammann-Beenker tiling, the
situation is simpler; the acceptance domain, which is obtained as the
projection of the four-dimensional hypercube to $E_{\perp}$, is a
regular octagon $O$.

Now, considering an arbitrary motive $c$ consisting of a collection of
$p$ points $c=\{\bi{r}_\parallel^{(i)}\, ; \; 1\!\le\! i\!\le\! p\}$
in physical space, we can compute its occurrence frequency, i.e., how
often translated copies of the point set occur in the infinite
tiling. Associated to the set $c$ of points in physical space is a
corresponding acceptance domain $A(c)\subset A$ in perpendicular
space, obtained by intersecting $p$ copies of the acceptance domain
$A$ shifted appropriately with respect to each other. This corresponds
to the acceptance domain filled by choosing a reference point of the
motive $c$, and, for all occurrences of the motive in an infinite
tiling, lifting the positions of this reference point to the
higher-dimensional lattice and projecting to $E_\perp$.  Hence, the
area of $A(c)$, divided by the area of $A$, is the occurrence
frequency of our motive, as follows from the uniform distribution on
the acceptance domain, see \cite{Hof} and references therein.

\begin{figure}
\centerline{\psfig{figure=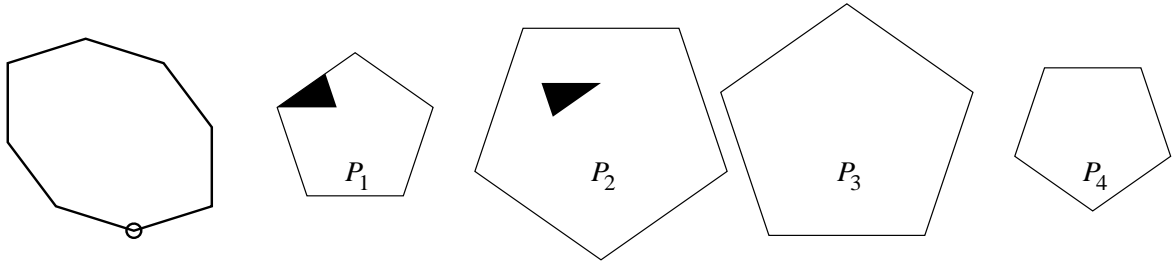,width=\textwidth}}
\caption{The ``fattest'' loop of length 8 in the Penrose lattice and
         its acceptance domain (black polygons) with respect to the
         reference point marked by the circle ($\circ$). The area
         fraction is $\tau-8/5\simeq 0.0180$ where
         $\tau=(1+\sqrt{5})/2$ is the golden ratio. The symmetry
         factors read $R=5$ and $S=1$, thus the occurrence
         frequency of this loop
         in the Penrose tiling, in an arbitrary orientation, is
         $5\tau-8\simeq 0.0902$.  \label{fig:ad1}}
\end{figure}

\begin{figure}
\centerline{\psfig{figure=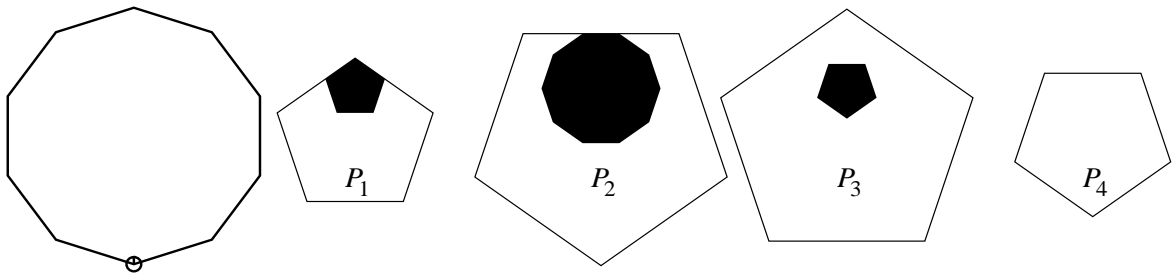,width=\textwidth}}
\caption{The same as \fref{fig:ad1}, now for the 
         ``fattest'' loop of length $10$. Here,  
         the area fraction is $(14\tau-22)/5\simeq 0.1305$, and
         the symmetry factors read $R=S=1$.\label{fig:ad2}}
\end{figure}

In the Penrose case, the acceptance domain $A(c)$ consists of four
pieces $A_{m}(c)\subset P_{m}$ ($m=1,2,3,4$) which have to be taken
into account. They are given by
\begin{equation} 
A_m(\{ \bi{r}_\parallel^{(i)} \}) \; =\; \bigcap\limits_i \left\{
P_{m + t^{(i)}} - \bi{r}_\perp^{(i)} \right\} 
\end{equation}
where $P_m=\emptyset$ if $m\not\in\{0,1,2,3,4,5\}$. The coordinates
$\bi{r}_{\parallel}^{(i)}\in E_{\parallel}$ and
$\bi{r}_{\perp}^{(i)}\in E_{\perp}$ have the form
\begin{equation}
\bi{r}_{\parallel}^{(i)} \; = \;
\sum\limits_{j=0}^4 n_j^{(i)}   \left( \begin{array}{@{}c@{}}
\cos{\frac{2 \pi j}{5}} \\ \sin{\frac{2 \pi j}{5}} \end{array} \right) \; ,
\qquad
\bi{r}_\perp^{(i)} \; = \;
\sum\limits_{j=0}^4 n_j^{(i)}  \left( \begin{array}{@{}c@{}}
1\\ \cos{\frac{4\pi j}{5}} \\ \sin{\frac{4\pi j}{5}} \end{array} \right) \; ,
\end{equation} 
with integer coefficients $n_{j}^{(i)}$ which correspond to the
coordinates of the lattice point in $\mathbb{Z}^5$ that projects to
$\bi{r}_\parallel^{(i)}$. The first component of $\bi{r}_{\perp}^{(i)}$,
\begin{equation}
t^{(i)} \; = \; \sum_{j=0}^4 n_j^{(i)} \; ,
\end{equation}
denotes the so-called translation class of the point
$\bi{r}_\parallel^{(i)}$, which just labels the part of the acceptance
domain $P_{t^{(i)}}$ where the corresponding perpendicular projection
lies. In figures \ref{fig:ad1} and \ref{fig:ad2}, we show two examples
where the motives are the ``fattest'' loops, in terms of the enclosed
area, of length $8$ and $10$ in the Penrose tiling that contribute to
the high-temperature expansion.

For the eight-fold Ammann-Beenker case there is only one acceptance
domain $O$, hence
\begin{equation}
A(\{ \bi{r}_\parallel^{(i)} \})
\; =\; \bigcap\limits_i \left\{ O-\bi{r}_\perp^{(i)}\right\} \; ,
\end{equation}
where the projections to $E_{\parallel}$ and $E_{\perp}$ are given by
\begin{equation}
\bi{r}_\parallel^{(i)} \; =
\; \sum\limits_{j=0}^3 n_j^{(i)}   \left( \begin{array}{@{}c@{}}
\cos{\frac{\pi j}{4}} \\ \sin{\frac{\pi j}{4}} \end{array} \right) \; ,
\qquad 
\bi{r}_\perp^{(i)} \; = \;
\sum\limits_{j=0}^3 n_j^{(i)}   \left(
\begin{array}{@{}c@{}} \cos{\frac{3\pi j}{4}} \\ \sin{\frac{3\pi j}{4}}
\end{array} \right) \; .
\end{equation} 
Here, $n_{j}^{(i)}\in\mathbb{Z}$ denote the coordinates of the lattice
point in $\mathbb{Z}^4$ that projects to $\bi{r}_\parallel^{(i)}$. 

The acceptance domains of a motive $c$ are intersections of convex
polygons and hence themselves polygonal, see figures~\ref{fig:ad1} and
\ref{fig:ad2}.  It is readily seen that the coordinates of the
vertices of the acceptance domains belong to certain extensions of the
field of rational numbers $\mathbb{Q}$. For the Penrose tiling, one
has to perform the calculation in the field
\begin{equation}
\mathbb{Q}(\tau,\sqrt{2+\tau}) \;=\; 
\left\{\left. a+b\sqrt{2+\tau}+c\tau+d\tau\sqrt{2+\tau} \;\;\right|\;
a,b,c,d\in\mathbb{Q}\right\}
\end{equation}
where $\tau = (1+\sqrt{5})/2$ is the golden ratio, satisfying the
quadratic equation $\tau^2 = \tau + 1$.  For the
Ammann-Beenker case, the corresponding number field is
\begin{equation}
\mathbb{Q}(\lambda) \;=\; \left\{ a+b\lambda \mid
a,b\in\mathbb{Q}\right\}
\end{equation}
where $\lambda=1+\sqrt{2}$ is the ``silver mean'' that is a solution
of the quadratic equation $\lambda^2 = 2\lambda +1$. Therefore, in
order to compute the occurrence frequency of a given motive $c$ in the
tiling ${\cal G}$, we have to determine the area of the acceptance
domain carrying out the calculation in the appropriate number
field. The averaged lattice constant $\langle c\, ;{\cal G}\rangle$ is
the occurrence frequency of $c$ summed over all possible orientations
of the motive. In these quasiperiodic tilings, the frequencies of
motives are independent of their orientation, hence we do not need to
calculate them separately, but just have to count how many
orientations of the motive occur in the tiling.

\begin{figure}[tb]
\centerline{\psfig{figure=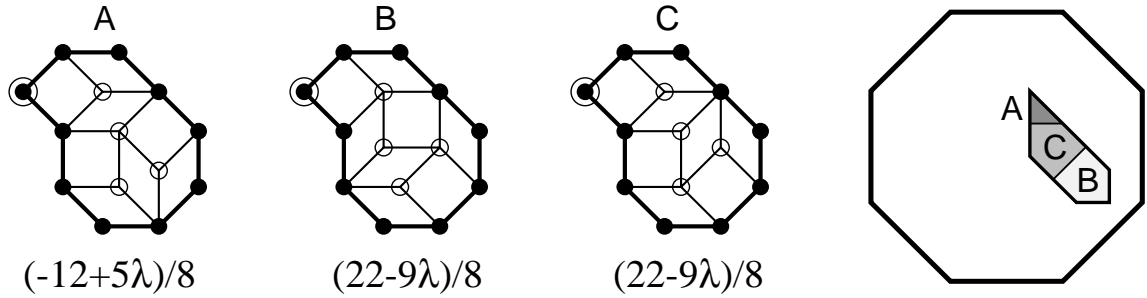,width=\textwidth,angle=180}} 
\caption{A loop of length 10 in the Ammann-Beenker tiling that can be
         filled in three different ways. The corresponding 
         occurrence frequencies
         of the filled patches, obtained from the area fraction of the
         acceptance domains shown on the right, are given below the patches. 
         They add up to $4-13\lambda/8\simeq 0.0769$ which is the
         frequency of the (empty) loop in the Ammann-Beenker
         tiling. The encircled node denotes the reference
         point.\label{fig:df1}}
\end{figure}

\begin{figure}[tb]
\centerline{\psfig{figure=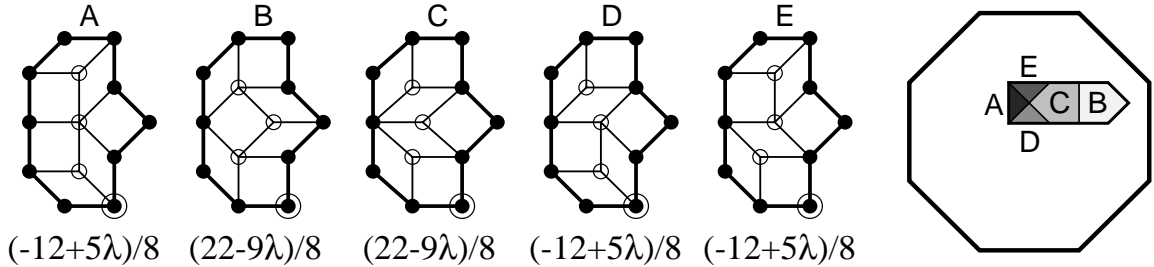,width=\textwidth,angle=180}}
\caption{Same as \fref{fig:df1} for another, reflection-symmetric loop
         of length $10$ which can be filled in five ways obtaining
         three reflection-symmetric patches and one pair of patches
         that map onto each other under reflection. Here, the
         frequency of the (empty) loop is $1-3\lambda/8\simeq
         0.0947$.\label{fig:df2}}
\end{figure}

Let us focus on the Penrose tiling as an example. Rotating the motive
$c$ by an angle $\pi k/5$ ($k\in\mathbb{Z}$) essentially corresponds
to a rotation of the acceptance domain by $2\pi k/5$. Furthermore,
also the mirror image $\bar{c}$ of the motive $c$ occurs with the same
frequency, since the corresponding acceptance domains $A_{m}(\bar{c})$
are just $-A_{5-m}(c)$. Therefore, in our expansion
\eref{eq:expansion2}, it is advantageous to jointly consider graphs
which are mirror images of each other because they give the same
contribution. For this reason, we assign two symmetry factors
$R\in\{1,2,5,10\}$ and $S\in\{1,2\}$ to each of these graphs, $R$
counting the number of rotations by angles $\pi k/5$ which do not map
the graph onto itself, and $S=2$ if reflection does not map the graph
onto itself or onto a rotated copy of itself, compare figures
\ref{fig:ad1} and \ref{fig:ad2}. The averaged lattice constant
$\langle c\, ;{\cal G}\rangle$, as defined above, is thus $R$ times
the area fraction obtained for a fixed orientation of the graph
$c$. Multiplying $\langle c\, ;{\cal G}\rangle$ by the factor $S$, we
can restrict the sum in equation \eref{eq:expansion2} to graphs that
are non-equivalent under reflection.

Eventually, we have to consider all star subgraphs of the
quasiperiodic tiling, corresponding to all possible fillings of loops.
In contrast to the case of simple planar (periodic) lattices, a loop
in the quasiperiodic tilings can have several fillings, which may
occur with different frequencies. In figures \ref{fig:df1} and
\ref{fig:df2}, the possible fillings, together with the corresponding
frequencies, of two exemplary loops in the Ammann-Beenker tiling are
shown. In orthogonal space, the different fillings correspond to a
dissection of the acceptance domain of the loop into non-overlapping
parts, see figures \ref{fig:df1} and \ref{fig:df2}. 

In order to avoid confusion, we would like to point out once more how
our frequencies are normalized, i.e., what the numbers given in
figures \ref{fig:df1} and \ref{fig:df2} really mean. We emphasize that
the frequency we compute is {\em not}\/ the frequency of a particular
loop of length $n$ among all loops of the same length. Instead, it
gives the probability that a randomly chosen vertex belongs to the
particular loop, in an arbitrary orientation.

\section{Expansion coefficients for the Penrose and the Ammann-Beenker 
tiling}
\label{sec4}

The Penrose and the Ammann-Beenker tiling are both bipartite graphs,
which means that all closed loops have an even number of edges, and at
least four.  Therefore, for zero magnetic field, only even powers of
$w$ occur in the expansion \eref{eq:expansion2} that takes the form
\begin{equation}
 F(w) \;=\; \lim_{N\rightarrow\infty}\frac{1}{N}\ln \tilde{Z}({\cal G}_N)
\; = \; \sum\limits_{n=2}^\infty g_{2n}\, w^{2n}
\label{eq:expansion}
\end{equation} 
where ${\cal G}_N$ denotes a finite patch of the quasiperiodic graph
${\cal G}$ containing $N$ vertices, and $F(w)$ is, apart from a factor
$-1/\beta$, the free energy per vertex. We calculated the expansion
coefficients $g_{2n}$ up to $18$th order in $w$ for both the Penrose
and the Ammann-Beenker tiling. The results are presented in
\tref{tab:ec}.

\begin{table}[bt]
\caption{The expansion coefficients $g_{2n}$ of the free energy 
         of the zero-field Ising model on the Penrose and the 
         Ammann-Beenker tiling. The values for the square lattice
         are included for comparison.\label{tab:ec}}
\begin{small}
\begin{tabular}{@{}rcr@{\;}c@{\;}rcr@{\;}c@{\;}rcr@{}l@{}} 
\br
 \centre{1}{$2n$} & \quad &
 \centre{3}{Penrose tiling} & \quad &
 \centre{3}{Ammann-Beenker tiling} & \quad &
 \centre{2}{Square lattice}\\
\mr
$4$   && $1$ &$=$& $1.00$  
          && $1$ &$=$& $1.00$ 
          && \qquad\qquad $1$ & \\
$6$   && $9 - 4\,\tau$ &$\simeq$& $2.53$
          && $\lambda$ &$\simeq$& $2.41$ 
          && $2$ & \\
$8$   && $12\case{1}{2} - 4\,\tau$ &$\simeq$& $6.03$ 
          && $47\case{1}{2}  - 17\,\lambda$ &$\simeq$& $6.46$
          && $4$&$\case{1}{2}$ \\
$10$  && $251\case{3}{5} - 144\case{1}{5}\,\tau$ &$\simeq$& $18.28$  
          && $138 - 50\,\lambda$ &$\simeq$& $17.29$ 
          && $12$ & \\
$12$  && $731\case{5}{6}  - 416\,\tau$ &$\simeq$& $58.73$ 
          && $803\case{1}{3} - 310\case{1}{2}\,\lambda$ &$\simeq$& $53.72$
          && $37$&$\case{1}{3}$ \\
$14$  && $1784 - 969\,\tau$ &$\simeq$& $216.13$
          && $-1220 + 586\,\lambda$ & $\simeq$& $194.73$ 
          && $130$ & \\
$16$  && $-27821\case{3}{4} + 17750\,\tau$ &$\simeq$& $898.35$
          && $96\case{3}{4} + 295\case{1}{2}\,\lambda$ &$\simeq$& $810.15$ 
          && $490$&$\case{1}{4}$\\
$18$  && $-124027 + 79078\case{2}{3}\,\tau$ &$\simeq$& $3924.97$
          && $-108706 + 46566\case{1}{3}\,\lambda$ &$\simeq$& $3715.07$ 
          && $1958$&$\case{2}{3}$ \\ 
\br  
\end{tabular}
\end{small}
\end{table}

As a by-product, we obtain information on another interesting physical
model, namely the problem of self-avoiding polygons, or closed
self-avoiding walks, on the quasiperiodic tiling. The quantities of
interest are the sums $S_{2n}$ of the occurrence frequencies of all
order-$2n$ loops which are presented in table~\ref{tab:cl}. Here,
$S_{2n}$ is nothing but the mean number per vertex of closed
self-avoiding walks with $2n$ steps, i.e., random walks with $2n$
steps that never return to a vertex visited before, except for the end
point which equals their starting point. For regular and recently also
for ``semi-regular'' lattices, there exist data for rather large
values of $n$ in the literature \cite{JenGut}; the square lattice
numbers are series M1780 in \cite{SloPlo}. A related problem, the
enumeration of self-avoiding walks on quasiperiodic tilings, was
already investigated by Briggs \cite{Briggs}. However, his results are
based on counting walks emanating from a fixed starting point, whereas
we compute the exact average over all possible starting points for the
self-avoiding polygons. Note that the number of walks does depend on
the initial vertex; however, the asymptotic behaviour should be
independent of this choice.

The coefficients $g_{2n}$ and $S_{2n}$ listed in tables \ref{tab:ec}
and \ref{tab:cl} belong to degree-$2$ extensions of the field of
rational numbers, namely $\mathbb{Q}(\tau)$ for the Penrose and
$\mathbb{Q}(\lambda)$ for the Ammann-Beenker tiling, respectively. We
note that for the Penrose case the frequencies of subgraphs, and thus
the coefficients $g_{2n}$ and $S_{2n}$, belong to the field
$\mathbb{Q}(\tau)$, whereas the areas of their acceptance domains in
general are elements of $\mathbb{Q}(\tau,\sqrt{2+\tau})$.

\begin{table}[tb]
\caption{The mean number (per vertex) of self-avoiding $2n$-step
         polygons $S_{2n}$ on the Penrose and the Ammann-Beenker tiling,
         and on the square lattice.\label{tab:cl}}
\begin{small}
\begin{indented}
\item[]\begin{tabular}{@{}rcr@{\;}c@{\;}rcr@{\;}c@{\;}r@{\;\;}c@{\;\;}r@{}} 
\br
\centre{1}{$2n$} &  & 
\centre{3}{Penrose tiling} &  &
\centre{3}{Ammann-Beenker tiling} & &
\centre{1}{Square lattice} \\ \mr
$4$      && $1$ &$=$& $1.00$
         && $1$ &$=$& $1.00$ 
         && $1$ \\
$6$      && $9 - 4\,\tau$ &$\simeq$& $2.53$
         && $\lambda$ &$\simeq$& $2.41$ 
         && $2$\\
$8$      && $15 - 4\,\tau$ &$\simeq$& $8.53$
         && $50  - 17\,\lambda$ &$\simeq$& $8.96$ 
         && $7$ \\
$10$     && $309\frac{3}{5} - 168\frac{1}{5}\,\tau$ &$\simeq$& $37.45$
         && $142 - 44\,\lambda$ &$\simeq$& $35.77$ 
         && $28$ \\
$12$     && $1066 - 552\,\tau$ &$\simeq$& $172.85$
         && $1173  - 416\,\lambda$ &$\simeq$& $168.69$ 
         && $124$ \\
$14$     && $6400 - 3405\,\tau$ &$\simeq$& $890.59$
         && $1704  - 353\,\lambda$ &$\simeq$& $851.78$ 
         && $588$ \\
$16$     && $5093 - 170\,\tau$ &$\simeq$& $4817.93$
         && $27175 - 9356\,\lambda$ &$\simeq$& $4587.62$ 
         && $2938$ \\
$18$     && $75115 - 29655\,\tau$ &$\simeq$& $27132.20$
         && $5992 + 8178\,\lambda$ &$\simeq$& $25735.44$ 
         && $15268$ \\
\br        
\end{tabular}
\end{indented}
\end{small}
\end{table}

The limitation of our calculations was caused by a strong, exponential
growth of the number of graphs which have to be taken into account.
For the Penrose tiling, we have --- even after identifying graphs that
are equivalent by rotation or reflection --- to deal with more than
$300\, 000$ different graphs contributing to the $18$th order, see
\tref{tab:q}, and their quantity grows approximately by a factor
between $6$ and $7$ when increasing the order by $2$. The
corresponding number of graphs for the square lattice, included in
\tref{tab:q}, are much smaller; the sequence of these numbers is
apparently not contained in \cite{SloPlo}. We generated the order-$2n$
loops as boundaries of patches that are constructed iteratively by
successively attaching rhombi to their surface, terminating the
process when attaching further rhombi does not lead to new order-$2n$
loops. By this procedure, we make sure that all graphs are
found. However, we have to pay the price that topologically identical
graphs are obtained repeatedly and have to be rejected, thus slowing
down the procedure substantially.

\begin{table}[h]
\caption{The number of symmetry-inequivalent closed loops of order $2n$ 
         contributing to the high-temperature expansion and the number 
         of patches obtained by filling the loops.\label{tab:q}}
\begin{indented}
\lineup
\item[]\begin{tabular}{@{}lcrcrcrcrcr@{}} \br
& \qquad & 
\centre{3}{Penrose} & \qquad &
\centre{3}{Ammann-Beenker} & \quad\;\; &
\centre{1}{Square lattice}\\ \ns\ns
& &
\crule{3} & &
\crule{3} & &
\crule{1}\\
\centre{1}{$2n$} & &
\centre{1}{empty} & &
\centre{1}{filled} & & 
\centre{1}{empty} & &
\centre{1}{filled} & &
\centre{1}{empty/filled}\\ 
\mr
$\0 4$  && $2$       && $2$       && $2$       && $2$       && $1$ \\      
$\0 6$  && $6$       && $6$       && $4$       && $4$       && $1$ \\
$\0 8$  && $24$      && $28$      && $17$      && $20$      && $3$ \\
$10$    && $143$     && $174$     && $77$      && $112$     && $6$ \\
$12$    && $839$     && $1034$    && $479$     && $743$     && $25$ \\
$14$    && $5634$    && $6957$    && $3007$    && $4981$    && $86$ \\
$16$    && $37677$   && $46712$   && $20175$   && $35063$   && $414$ \\
$18$    && $255658$  && $317028$  && $139146$  && $244638$  && $1975$ \\
\br
\end{tabular}
\end{indented}
\end{table}

\section{Critical behaviour}
\label{sec5}

In many cases, high-temperature expansions yield good estimates of the
critical temperature and the critical exponent of the free energy. The
simplest approach, which is commonly used for this purpose, uses the
ratio of two successive coefficients $g_{2n}/g_{2n-2}$ in the
expansion \cite{Domb}. Assuming that the free energy $F(w)$ behaves in
the vicinity of the critical point $w_c$ as
\begin{equation}
 F(w)\;\sim\; \left(1 - w^2/w_c^2\right)^{\kappa}\; ,
\label{eq:crit_point}
\end{equation}
one can easily see from \eref{eq:expansion} that
\begin{equation} 
\frac{g_{2n}}{g_{2n-2}} \; =\; 
\frac{1}{w_c^2} \left( 1 - \frac{\kappa + 1}{n} \right) + \Or (n^{-2}) \; .
\label{eq:quot} 
\end{equation}
In other words, for sufficiently large values of $n$, the ratios
$g_{2n}/g_{2n-2}$ should lie on a straight line when plotted as a
function of $n^{-1}$. The slope of this line and its displacement
from the origin determine the critical point $w_c$ and the exponent
$\kappa$.

We may estimate the critical temperature from the sequence
\begin{equation}
\varrho (2n) = \left[n \frac{g_{2n}}{g_{2n-2}} - 
(n-1) \frac{g_{2n-2}}{g_{2n-4}}\right]^{-1} 
\end{equation}
that approaches $w_c^2$ in the limit $n\rightarrow \infty$. In
\tref{tab:aa}, we show the results for $\varrho (2n)$ for the two
quasiperiodic tilings under consideration and compare these with the
estimates of the critical point from Monte-Carlo simulations
\cite{OkaNii1,Sorensen,Ledue}. The corresponding values for the square
lattice are included for comparison.

\begin{table}[h]
\caption{Estimates of the critical point of the Ising model on the 
         Penrose tiling and the Ammann-Beenker tiling, and on the square 
         lattice.\label{tab:aa}}
\begin{indented}
\item[]\begin{tabular}{rclclcl} 
\br
& &
\centre{5}{$\varrho (2n)$} \\ \ns\ns
& &
\crule{5} \\ \ns
& \qquad & 
\centre{1}{Penrose} & \qquad &
\centre{1}{Ammann-Beenker} & \qquad & 
\centre{1}{Square}\\ \ns
$2n$ & &
\centre{1}{tiling} & & 
\centre{1}{tiling} & &
\centre{1}{lattice} \\
\mr
 $8$ && $0.5116$ && \quad $0.2892$ && $0.3333$ \\
$10$ && $0.1778$ && \quad $0.3725$ && $0.2308$ \\
$12$ && $0.2430$ && \quad $0.1902$ && $0.1875$ \\
$14$ && $0.1543$ && \quad $0.1486$ && $0.1752$ \\
$16$ && $0.1334$ && \quad $0.1264$ && $0.1726$ \\
$18$ && $0.1648$ && \quad $0.1252$ && $0.1728$ \\
\mr
\centre{1}{$w_c^2$} && 
$0.1563(5)^{\rm a}$ && 
\quad  $0.1566(5)^{\rm c}$ && 
$0.1716^{\rm d}$ \\
&& $0.1552(6)^{\rm b}$ && && \\
\br
\end{tabular}
\item[] $^{\rm a}$ After reference~\cite{OkaNii1}.
\item[] $^{\rm b}$ After reference~\cite{Sorensen}.
\item[] $^{\rm c}$ After reference~\cite{Ledue}.
\item[] $^{\rm d}$ This corresponds to the exact value $w_c=\sqrt{2}-1$ 
\cite{KraWan,Onsager}.
\end{indented}
\end{table} 

As one can see, the convergence of $\varrho (2n)$ is rather poor for
the quasiperiodic tilings. In general, the rate of convergence is
determined by additional singularities $w_{c}^{\prime}\in\mathbb{C}$
of $F(w)$ lying close to $w_c$ in the complex plane. These give a
correction to $g_{2n}/g_{2n-2}$ which behaves like $\Or
[(w_{c}^{\prime}/w_{c})^{2n}]$ \cite{Domb}. The influence of these
corrections must be substantial in our case rendering the method
rather inapplicable for us. We will come back to this point in
\sref{sec6} below when we discuss the corresponding quantities for
periodic approximants, compare also \fref{fig:app-ratios} that
contains a plot of the ratios $g_{2n}/g_{2n-2}$ for the case of the
Penrose tiling.

\begin{figure}[tb]
\psfig{figure=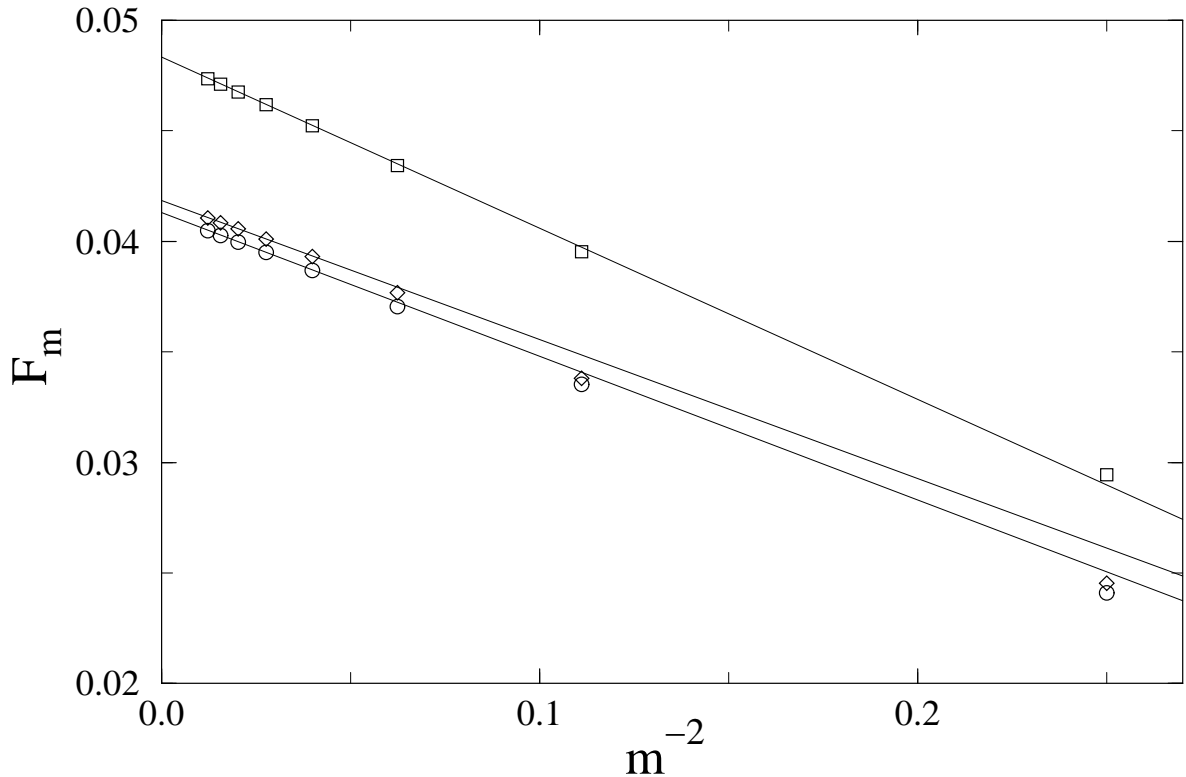,width=\columnwidth}
\caption{The dependence of the partial sums $F_m$ of 
         equation~\protect\eref{eq:fc}
         on $m^{-2}$ for the Penrose tiling ($\opencircle$), 
         the Ammann-Beenker tiling ($\opendiamond$),
         and the square lattice ($\opensquare$), respectively. 
         The straight lines are least-square fits to the 
         data, disregarding the three points with smallest 
         $m$ values.\label{fig:partial_sums}}
\end{figure}

There is, however, another method which is more suitable for us to
examine the critical behaviour.  Let us consider a sequence of partial
sums $F_m$ of the expansion \eref{eq:expansion} at the critical
point $w_c$
\begin{equation} 
F_m  \;=\; \sum\limits_{n=2}^{m} g_{2n} w_c^{2n} \; .
\label{eq:fc}
\end{equation}
If the function $F(w)$ behaves like \eref{eq:crit_point}, then the
asymptotic behaviour of the coefficient $\tilde{g}_{2n}=w_c^{2n}
g_{2n}$ of its expansion in the variable $w^2/w_c^2$ is given by
$\tilde{g}_{2n} \sim n^{-\kappa -1}$ for $n\rightarrow \infty$
\cite{Abe1,AbeDotOga}.  Therefore, for large $m$, we have
\begin{eqnarray}
F_m & = &     
F_{\infty} - \sum\limits_{n=m+1}^\infty 
g_{2n} w_{c}^{2n} \; = \; 
F_{\infty} - \sum\limits_{n=m+1}^\infty 
\tilde{g}_{2n} \nonumber\\
& \simeq &
F_{\infty} - \tilde{b}\sum\limits_{n=m+1}^\infty n^{-(\kappa +1)} 
\; \simeq\; F_{\infty} - bm^{-\kappa} 
\end{eqnarray}
where $b$ is a parameter and the last relation is obtained by
approximating the sum by an integral.  Therefore, for sufficiently
large $m$, the values $F_m$ should lie on a straight line when plotted
versus $m^{-\kappa}$.  In \fref{fig:partial_sums}, we plot the partial
sums $F_m$ for the Penrose and the Ammann-Beenker tiling, taking
$\kappa=2$ and $w_c$ equal to the Monte-Carlo estimates of
\cite{Sorensen,Ledue}, see also \tref{tab:aa}. For comparison, we also
included corresponding data for the square lattice where the exact
solution is known. Apparently, the data points lie close to a straight
line for all three cases, and the fluctuations in the data for the
quasiperiodic tilings are not visibly larger than those for the square
lattice.

\begin{figure}[tb]
\psfig{figure=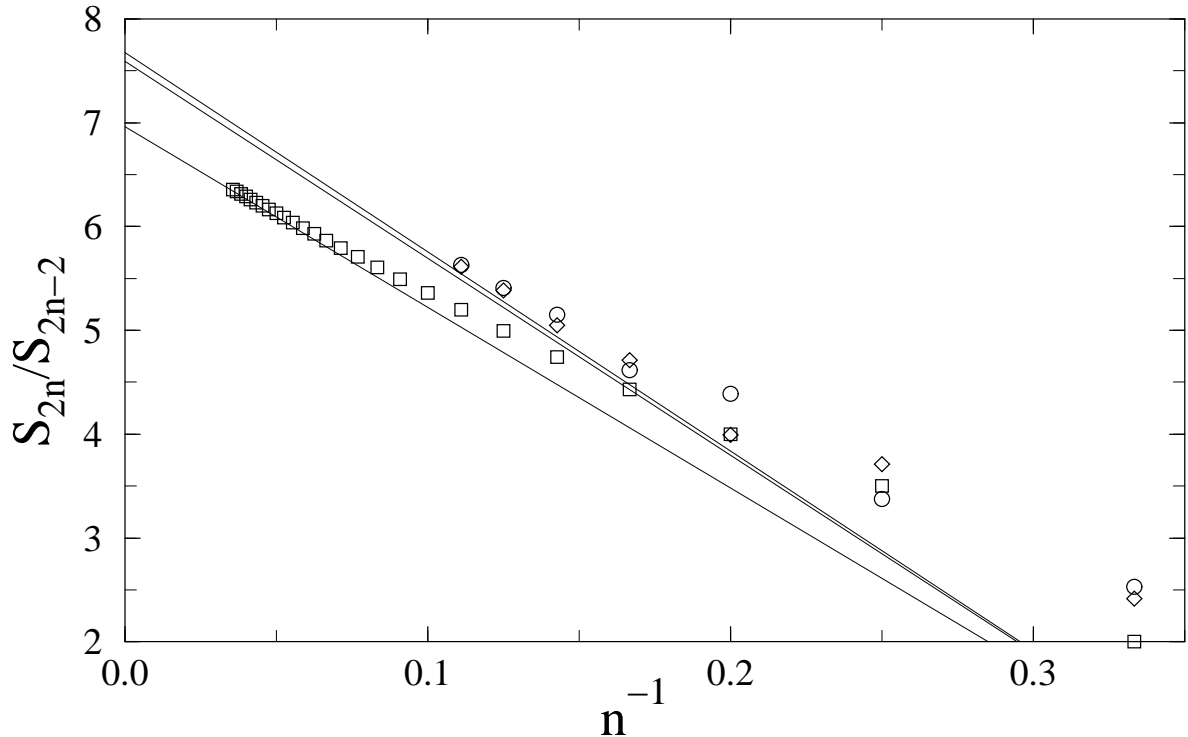,width=\columnwidth}
\caption{Ratios of the numbers of self-avoiding polygons (per vertex)
         on the Penrose tiling ($\opencircle$), 
         the Ammann-Beenker tiling ($\opendiamond$),
         and the square lattice ($\opensquare$), 
         respectively. Square lattice data for $2n\le 56$ are 
         taken from \cite{Guttmann}. The straight lines are
         obtained from equation \eref{eq:genfun}, in analogy to 
         equation \eref{eq:quot},
         using the critical exponent $\alpha=1/2$ and the 
         approximate values of the critical point $x_c$ given in 
         \cite{Briggs} and \cite{JenGut}.\label{fig:sap-ratios}}
\end{figure}

Concerning the numbers of self-avoiding polygons $S_{2n}$ shown in 
\tref{tab:cl}, one considers their generating function
\begin{equation}
G(x) \; =\; \sum_{n=2}^{\infty} S_{2n} x^{2n} 
\end{equation}
which has a critical point $x_c$ that is characterized by a cusp-like
singularity; i.e., in the vicinity of $x_c$ one has
\begin{equation}
G(x) \; \sim\; A(x) + B(x) \left(1-x^2/x^2_c\right)^{2-\alpha}
\label{eq:genfun}
\end{equation}
with a critical exponent $\alpha$, and $A(x)$ and $B(x)$ are
non-singular at $x=x_c$. We note that the only exact result for the
related problem of self-avoiding walks in two dimensions is obtained
by the Coulomb gas approach \cite{Nienhuis} and gives a critical point
$x_c^2=1/(2+\sqrt{2})$ and critical exponents $\alpha=1/2$,
$\gamma=43/32=1.34375$ and $\nu=3/4$ for the hexagonal
lattice. Frequently, the so-called connective constant $\mu=1/x_c$ is
given instead of $x_c$. In \cite{Briggs}, estimates of the critical
point $x_c$ for self-avoiding walks, which coincides with
the value for self-avoiding polygons, are given based on enumerations
of walks of at most $20$ and $16$ steps for the Penrose and the
Ammann-Beenker tiling, respectively. The corresponding critical
exponent in this case is $\gamma$, and all results support the
conjecture that the self-avoiding walk problems on two-dimensional
lattices and quasiperiodic tilings belong to the same universality
class.  

In \fref{fig:sap-ratios}, we show the ratios of successive numbers
$S_{2n}/S_{2n-2}$ as a function of $1/n$, which, by the same arguments
that led to equation \eref{eq:quot}, should lie on a straight line for
large $n$. Clearly, this is true for the square lattice, whereas the
data for the Penrose and the Ammann-Beenker tiling still show sizable
fluctuations. The straight lines in \fref{fig:sap-ratios} are the
functions $[1-5/(2n)]/x_c^2$, compare \eref{eq:quot}, where we used
the critical exponent $\alpha=1/2$ and the value $\mu=2.618\, 158\,
53$ cited in \cite{JenGut} for the square lattice connective constant,
and the estimates $x_c=0.363$ and $x_c=0.361$ \cite{Briggs} for the
critical points on the Penrose and the Ammann-Beenker tiling,
respectively. Given the rather short sequence at our disposal, and the
uncertainty in the estimates \cite{Briggs}, the agreement for the
quasiperiodic cases is reasonable, thus supporting the conjecture that
the critical point of self-avoiding polygons on such quasiperiodic
tilings is described by the same critical exponents as for the
hexagonal lattice \cite{Nienhuis}.

\section{Partition functions of periodic approximants}
\label{sec6}  

One may pose the question whether one can calculate the expansion
coefficients $g_{2n}$ in equation \eref{eq:expansion} by a different
method, thus verifying our results.  Perhaps it might even be possible
to calculate the partition function $Z(\cal G)$ on certain
quasiperiodic tilings $\cal G$ exactly. Although this may seem
hopeless, there exist methods to tackle this problem, which at least
allow us to compute the partition function of general periodic
lattices explicitly, thus also for periodic approximants of the
quasiperiodic tilings. Let us briefly recall some exact results on
partition functions on two-dimensional lattices.

The first solution of the two-dimensional zero-field Ising model for
the square lattice had been found by Onsager and Kaufman
\cite{Onsager,Kaufman} in 1944. Several years later, Kac and Ward
\cite{KacWard} developed a combinatorial approach in which the problem
was reduced to the calculation of a determinant of a certain matrix
$K$ (see below) which depends on the lattice and the coupling
constants between the spins.  Although this approach was not rigorous,
it appeared extremely plausible and it initiated numerous attempts to
generalize this result to other lattices
\cite{PottsWard,Sherman,Burgoyne,Vdo1}.  Recently, Dolbilin \etal
\cite{Dolbilin} proved the long-known formula
\begin{equation}
\tilde{Z}({\cal G})^2 \,=\, \det(K)
\label{eq:KacWardformula}
\end{equation}
for a zero-field Ising model on an arbitrary planar graph $\cal G$
with arbitrary (in general site-dependent) spin coupling constants.
The matrix elements $K(\bi{e}_{i},\bi{e}_{j})$ of the $2M\times 2M$
matrix $K$ are labeled by oriented edges $\bi{e}_i$ and $\bi{e}_j$, $1
\le i,j \le 2 M$. They are defined as
\begin{equation}
\fl
K(\bi{e}_{i}, \bi{e}_{j}) \; = \; \cases{
   1 & if $\bi{e}_i = \bi{e}_j$ \\
   -w_i \exp{[\frac{\rmi}{2} (\widehat{\bi{e}_i,\bi{e}_j}) ]} & 
if $f(\bi{e}_i) = b(\bi{e}_j)$ and  
      $f(\bi{e}_j) \ne b(\bi{e}_i)$ \\
   0 & otherwise\\ }
\label{eq:matrix}
\end{equation}
where $w_i=\tanh(\beta J_i)$ and $J_i$ is the spin coupling constant 
along the edge $\bi{e}_i$, which is independent of the edge in our
case, $J_i\equiv J$. Furthermore, 
$(\widehat{\bi{e}_{i},\bi{e}_{j}})$ denotes the angle between
edges $\bi{e}_i$ and $\bi{e}_j$, and $b(\bi{e}_i)$ and
$f(\bi{e}_i)$ are the starting point and the end point of the
edge $\bi{e}_i$, respectively.  If $\cal G$ is periodic, the
matrix $K$ is cyclic and the determinant can be calculated exactly in
the thermodynamic limit $M \rightarrow \infty$.  We can therefore apply
\eref{eq:matrix} and calculate $\tilde{Z}(\cal G)$ exactly for periodic
approximants of the Penrose and the Ammann-Beenker tiling.

Let us now briefly describe how to generate periodic approximants of
quasiperiodic tilings in the framework of the cut-and-project method
discussed in \sref{sec3}. The acceptance domain $A$ and the projection
onto perpendicular space $E_{\perp}$ are altered in a way that
corresponds to replacing the irrational numbers $\tau$ and $\lambda$
by rational approximants $\tau_m$ and $\lambda_m$. Here, for the
Penrose tiling we use $\tau_m=f_{m+1}/f_m$ where
$f_{m+1}=f_{m}+f_{m-1}$, and $f_{0}=0$, $f_{1}=1$ are the Fibonacci
numbers, and $\lim_{m\rightarrow\infty}\tau_m=\tau$. Analogously, one
defines rational approximants $\lambda_m=g_{m+1}/g_{m}$ with the
``octonacci numbers'' $g_{m+1}=2g_{m}+g_{m-1}$ and $g_{0}=0$,
$g_{1}=1$, and $\lim_{m\rightarrow\infty}\lambda_m=\lambda$ for the
case of the Ammann-Beenker tiling.

\begin{table}[tb]
\caption{Expansion coefficients of the free energy for the Ising model
on the Penrose tiling and its periodic approximants $m=1,2,3,4,5,6$
with ${\cal N}$ vertices in a unit cell.\label{tab:expPenrose}}
\begin{small}
\begin{tabular}{@{}ccrcrcrcr@{$\;\simeq\;$}rcr@{$\;\simeq\;$}r@{}} 
\br
$m$ &\mbox{\quad\;\,}& 
${\cal N}$ &\mbox{\quad\;\,}& 
$2M$ &\mbox{\quad\;\,}& 
\centre{1}{$g_{4}$} &\mbox{\quad\;\,}& 
\centre{2}{$g_{6}$} &\mbox{\quad\;\,}& 
\centre{2}{$g_{8}$}\\
\mr
$1$ &&   $11$ &&   $44$ && $1$ && $25/11$    & $2.2727$ 
    && $127/22$    & $5.7727$ \\
$2$ &&   $29$ &&  $116$ && $1$ && $73/29$    & $2.5172$ 
    && $349/58$    & $6.0172$ \\
$3$ &&   $76$ &&  $304$ && $1$ && $5/2$      & $2.5000$ 
    && $227/38$    & $5.9737$ \\
$4$ &&  $199$ &&  $796$ && $1$ && $503/199$  & $2.5276$ 
    && $2399/398$  & $6.0276$ \\
$5$ &&  $521$ && $2084$ && $1$ && $1315/521$ & $2.5240$ 
    && $6273/1042$ & $6.0202$ \\
$6$ && $1364$ && $5456$ && $1$ && $862/341$  & $2.5279$ 
    && $4111/682$  & $6.0279$ \\
\mr
$\infty$ && $\infty$ && $\infty$ 
&& $1$ 
&& $9-4\tau$ & $2.5279$ 
&& $12\frac{1}{2}-4\tau$ & $6.0279$ \\
\br
\end{tabular}
\begin{tabular}{@{}ccr@{$\;\simeq\;$}rcr@{$\;\simeq\;$}rcr@{$\;\simeq\;$}r@{}} 
\br
$m$ &\mbox{\quad}& 
\centre{2}{$g_{10}$} &\mbox{\quad\,}& 
\centre{2}{$g_{12}$} &\mbox{\quad\,}& 
\centre{2}{$g_{14}$} \\
\mr
$1$      && $175/11$     & $15.909$ && $3145/66$      & $47.652$ 
         && $1812/11$     & $164.73$ \\
$2$      && $504/29$     & $17.379$ && $341/6$        & $56.833$ 
         && $6011/29$     & $207.27$ \\
$3$      && $679/38$     & $17.868$ && $6629/114$     & $58.149$ 
         && $16123/76$    & $212.14$ \\
$4$      && $3624/199$   & $18.211$ && $69833/1194$   & $58.487$ 
         && $42552/199$   & $213.83$ \\
$5$      && $9496/521$   & $18.226$ && $610783/10420$ & $58.616$ 
         && $112451/521$  & $215.84$ \\
$6$      && $24921/1364$ & $18.271$ && $160129/2728$  & $58.698$ 
         && $294347/1364$ & $215.80$ \\
\mr
$\infty$ && 251$\frac{3}{5}-144\frac{1}{5}\tau$ & $18.279$
         && $731\frac{5}{6}-416\tau$ & $58.731$ 
         && $1784-969\tau$ & $216.13$ \\
\br
\end{tabular}
\begin{tabular}{@{}ccr@{$\;\simeq\;$}rcr@{$\;\simeq\;$}r@{}} 
\br
$m$ &\mbox{\qquad}& \centre{2}{$g_{16}$} 
    &\mbox{\qquad}& \centre{2}{$g_{18}$} \\
\mr
$1$     && $29439/44$     & $669.07$ && $95119/33$     & $2882.4$ \\
$2$     && $100769/116$   & $868.70$ && $342484/87$    & $3936.6$ \\
$3$     && $33325/38$     & $876.97$ && $222817/57$    & $3909.1$ \\
$4$     && $709087/796$   & $890.81$ && $2345981/597$  & $3929.6$ \\
$5$     && $1867989/2084$ & $896.35$ && $6128605/1563$ & $3921.1$ \\
$6$     && $1223683/1364$ & $897.13$ && $4015369/1023$ & $3925.1$ \\
\mr
$\infty$ && $-27821\frac{3}{4}+17750\tau$ & $898.35$
         && $-124027+79078\frac{2}{3}\tau$ & $3925.0$ \\
\br
\end{tabular}
\end{small}
\end{table}

\begin{table}[tb]
\caption{Same as table~\protect\ref{tab:expPenrose},
         but for approximants $m=1,2,3,4$ of 
         the Ammann-Beenker tiling.\label{tab:expAmmann}}
\begin{small}
\begin{tabular}{@{}ccrcrcrcr@{$\;\simeq\;$}rcr@{$\;\simeq\;$}r@{}} 
\br
$m$ &\mbox{\quad\,}& 
${\cal N}$ &\mbox{\quad\,}& 
$2M$ &\mbox{\quad\;\;}& 
\centre{1}{$g_{4}$} &\mbox{\quad\,}& 
\centre{2}{$g_{6}$} &\mbox{\quad\,}& 
\centre{2}{$g_{8}$} \\
\mr
$1$ &&    $7$ &&   $28$ && $1$ && $17/7$       & $2.4286$ 
                               && $87/14$      & $6.2143$ \\
$2$ &&   $41$ &&  $164$ && $1$ && $99/41$      & $2.4146$
                               && $529/82$     & $6.4512$ \\
$3$ &&  $239$ &&  $956$ && $1$ && $577/239$    & $2.4142$
                               && $3087/478$   & $6.4582$ \\
$4$ && $1393$ && $5572$ && $1$ && $3363/1393$  & $2.4142$
                               && $17993/2786$ & $6.4584$ \\
\mr
$\infty$ && $\infty$ && $\infty$ 
&& $1$ 
&& $\lambda$ & $2.4142$ 
&& $47\frac{1}{2}-17\lambda$ & $6.4584$ \\
\br
\end{tabular}
\begin{tabular}{@{}ccr@{$\;\simeq\;$}rcr@{$\;\simeq\;$}rcr@{$\;\simeq\;$}r@{}} 
\br
$m$ &\mbox{\quad\,}& 
\centre{2}{$g_{10}$} &\mbox{\quad\,}& 
\centre{2}{$g_{12}$} &\mbox{\quad\,}& 
\centre{2}{$g_{14}$} \\
\mr
$1$      && $115/7$       & $16.429$ && $2189/42$    & $52.119$
         && $1395/7$      & $199.29$ \\
$2$      && $708/41$      & $17.268$ && $13183/246$  & $53.589$
         && $7994/41$     & $194.97$ \\
$3$      && $4132/239$    & $17.289$ && $77029/1434$ & $53.716$ 
         && $46542/239$   & $194.74$ \\
$4$      && $24084/1393$  & $17.289$ && $74832/1393$ & $53.720$
         && $271258/1393$ & $194.73$ \\
\mr
$\infty$ && $138-50\lambda$ & $17.289$
         && $803\frac{1}{3}-310\frac{1}{2}\lambda$ & $53.720$ 
         && $-1220+586\lambda$ & $194.73$\\
\br
\end{tabular}
\begin{tabular}{@{}ccr@{$\;\simeq\;$}rcr@{$\;\simeq\;$}r@{}} 
\br
$m$ &\mbox{\qquad}& 
\centre{2}{$g_{16}$} &\mbox{\qquad}& 
\centre{2}{$g_{18}$} \\
\mr
$1$     &&   $22815/28$   & $814.82$ && $78479/21$    & $3737.1$ \\
$2$     && $132885/164$   & $810.27$ && $153121/41$   & $3734.7$ \\
$3$     && $774507/956$   & $810.15$ && $2664121/717$ & $3715.6$ \\
$4$     && $4514157/5572$ & $810.15$ && $739303/199$  & $3715.1$ \\
\mr
$\infty$ && $96\frac{3}{4}+295\frac{1}{2}\lambda$ & $810.15$
         && $-108706+46566\frac{1}{3}\lambda$ & $3715.1$\\
\br
\end{tabular}
\end{small}
\end{table}

\begin{figure}[tb]
\psfig{figure=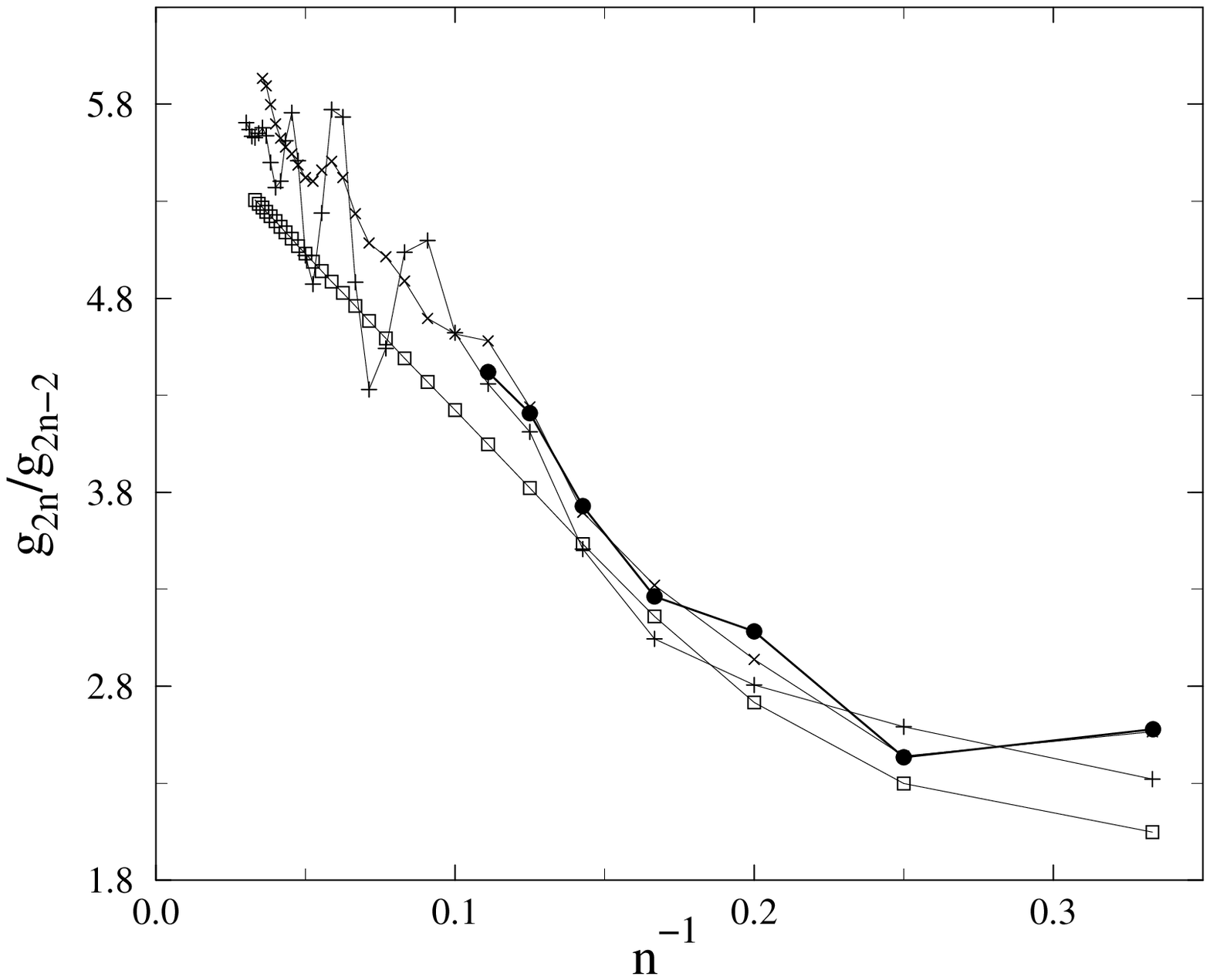,width=\columnwidth}
\caption{The ratios $g_{2n}/g_{2n-2}$ of expansion coefficients for
         the square lattice ($\opensquare$), the first ($+$) and the second 
         ($\times$) periodic approximant of the Penrose tiling, 
         and for the Penrose tiling ($\fullcircle$),
         respectively. Lines are meant as guides to the eye 
        only.\label{fig:app-ratios}}
\end{figure}

In this way, one obtains periodic approximants of the Penrose tiling
with unit cells containing ${\cal N}=11,29,76,199,521,1364$ vertices
for $m=1,2,3,4,5,6$, respectively.  The unit cells of the periodic
approximants of the Ammann-Beenker tiling with $m=1,2,3,4$ contain
${\cal N}=7,41,239,1393$ vertices. For both tilings, the number of
oriented edges is $2M=4{\cal N}$, because each vertex has exactly four
neighbours.  We note that the approximant $m+1$ contains about
$\tau^{2}=\tau+1\simeq 2.618$ and $\lambda^{2}=2\lambda+1\simeq 5.828$
as many vertices and bonds as the approximant $m$ for the Penrose and
the Ammann-Beenker case, respectively.

For each periodic approximant, we define a matrix $\tilde{K}$ labeled
by oriented edges $\bi{e}_i^{\prime}$ and $\bi{e}_j^{\prime}$ with
starting point in the unit cell. It is related to the Kac-Ward matrix
$K$ \eref{eq:matrix} as
\begin{equation}
\tilde{K}(\bi{e}_i^{\prime},\bi{e}_j^{\prime}) \; :=\; 
K(\bi{e}_i,\bi{e}_j) 
\exp{(-\rmi k_1 \Delta_1)} \exp{(-\rmi k_2 \Delta_2)} \; .
\end{equation}
Here, we can assume that $\bi{e}_i^{\prime}=\bi{e}_i$ starts in the
unit cell, and $\bi{e}_j^{\prime}$ equals $\bi{e}_j$ modulo the unit
cell, i.e., if $b(\bi{e}_j)=\xi_j V +\eta_j W$, where $V$ and $W$ are
the base vectors spanning the unit cell, then $b(\bi{e}_j^{\prime})=
\mbox{frac($\xi_j$)} V + \mbox{frac($\eta_j$)} W$ where frac($x$)
denotes the fractional part of $x$.  The integers $\Delta_1$ and
$\Delta_2$ are the integer parts $\lfloor\xi_j\rfloor$ and
$\lfloor\eta_j\rfloor$, respectively, and $\tilde{K}$ depends on the
``wave vectors'' $k_1$ and $k_2$ which are real numbers. Due to the
fact that the Kac-Ward matrix $K$ of the periodic approximant is
cyclic, its determinant can be expressed as a product of the
determinant of $\tilde{K}$ over all values of $k_1$ and $k_2$,
corresponding to a reduction to the unit cell.  Calculating the
logarithm of $\tilde{Z}({\cal G})$, and taking relation
\eref{eq:KacWardformula} into account, one obtains
\begin{equation}
\ln{\tilde{Z}({\cal G})} \,=\, \frac{1}{8 \pi^2} \int\limits_0^{2 \pi} 
\int\limits_0^{2 \pi} \; 
\ln{\mbox{det}\tilde{K}(k_1,k_2)}\; \rmd k_1 \rmd k_2 \; .
\label{eq:freeenergy}
\end{equation}
Let us now expand this equation in a series with respect to $w$ and
compare it with the high-temperature expansion \eref{eq:expansion}.
For this purpose, we exploit the fact \eref{eq:matrix} that the 
(finite-dimensional) matrix $\tilde{K}(k_1,k_2)$ has a form
$\tilde{K}(k_1,k_2)=1+w\tilde{L}(k_1,k_2)$ where $\tilde{L}(k_1,k_2)$
has zero trace.  Therefore, using
\begin{eqnarray}
\det{[ 1 + w \tilde{L}(k_1,k_2) ]} &=&
\det{\exp{\ln{[1 + w \tilde{L}(k_1,k_2)]}}} \nonumber \\
 & =& \exp{\tr\ln{[1 + w \tilde{L}(k_1,k_2)]}}
\end{eqnarray}
one obtains, expanding the logarithm in powers of $w\tilde{L}(k_1,k_2)$,
\begin{eqnarray}
\ln{\det{\tilde{K}(k_1,k_2)}} & =&  
\tr\ln{[1 + w \tilde{L}(k_1,k_2)]} \nonumber \\ & = &
\sum\limits_{p=1}^\infty \frac{(-1)^{p+1}}{p} 
\tr [\tilde{L}^{p}(k_1,k_2)] w^p \; ,
\end{eqnarray}
where, again, only even values of $p$ yield non-vanishing contributions
to the sum.
Comparing this result with equation \eref{eq:freeenergy}, we derive an
expression for the coefficient $g_{2n}$ in the expansion \eref{eq:expansion}
\begin{equation}
g_{2n} \;=\; -\frac{1}{16 \pi^2 n} \int\limits_0^{2 \pi} 
\int\limits_0^{2 \pi} \tr [\tilde{L}^{2 n}(k_1,k_2)] \; \rmd k_1 \rmd k_2
\label{eq:expansionKacWard}
\end{equation}
for the periodic approximants.  We have calculated the coefficients
from \eref{eq:expansionKacWard} for the leading orders in $w$ for both
the Penrose and the Ammann-Beenker tiling.  The limitation of the
calculation was due to a rapidly growing dimension of the complex
matrix $\tilde{L}(k_1,k_2)$, which was equal to $5456$ and $5572$ for
our largest approximants of the Penrose and Ammann-Beenker tiling,
respectively.  The results are presented in tables
\ref{tab:expPenrose} and \ref{tab:expAmmann}. Clearly, with increasing
size of the approximant, the coefficients approach those of the
quasiperiodic system, and the coefficients of the largest approximant
are already quite close to those of the quasiperiodic case.

We now consider the ratios \eref{eq:quot} for the periodic
approximants of the Penrose tiling. The result is shown in
\fref{fig:app-ratios}. Although we included terms up to order $2n=56$,
the data for the two periodic approximants do not lie on straight
lines, in contrast to those of the square lattice. Instead, they show
large fluctuations, and apparently the fluctuations for the smallest
approximant with $11$ vertices in the unit cell turn out to be much
larger than those for the larger approximant which contains $29$
vertices. It would be interesting to have a better understanding of this
phenomenon, perhaps an investigation of the complex-temperature phase
diagram of the periodic approximants can give an explanation of this
observation.  Again, \fref{fig:app-ratios} also shows that the data
for the second approximant are already rather close to that of the
quasiperiodic tiling, and one might conclude from \fref{fig:app-ratios}
that the fluctuations in the ratios of expansion coefficients become
less with increasing size of the approximant.

\section{Conclusions}
\label{sec7}  

We considered the Ising model on two planar quasiperiodic graphs, the
Penrose and the Ammann-Beenker tiling. We calculated the leading terms
of the high-temperature expansion of the free energy exactly, using
the embedding of the quasiperiodic tilings into higher-dimensional
periodic lattices to compute the occurrence frequencies of patterns in
the tiling. These frequencies are expressed in terms of characteristic
quadratic irrationalities related to ten- and eightfold rotational
symmetry, the golden mean $\tau=(1+\sqrt{5})/2$ and the silver mean
$\lambda=1+\sqrt{2}$ for the Penrose and the Ammann-Beenker tiling,
respectively.

The number of graphs that contribute to a given order in the expansion
grows much faster with the order than for a simple periodic lattice,
therefore we did not go beyond the 18th order in the expansion
variable $w=\tanh{(\beta J)}$ in this work. {}From our expansion
alone, it is difficult to extract information about the critical
behaviour. However, using estimates of the critical temperature
obtained by other methods to analyze our data, we find that our
expansions are in accordance with the the conjecture that Ising models
on such planar quasiperiodic graphs belong to the Onsager universality
class. 

In order to compute the expansion coefficients, we had to construct
all polygons on the quasiperiodic graphs with up to $2n=18$
edges. Thus, we obtain the average number of such self-avoiding
polygons as a by-product of our calculation. Comparison with earlier
results on self-avoiding walks \cite{Briggs}, based on enumerating
walks that start from a chosen vertex in the tiling, indicates that
the self-avoiding polygons on the examples we considered belong to
the same universality class. In particular, this means that the
corresponding critical point is described by the same exponents as
for the hexagonal lattice which are known analytically \cite{Nienhuis}.

Finally, we considered periodic approximants of the quasiperiodic
tilings. For these it is, in principle, possible to compute the free
energy of the infinite periodic system analytically. Here, we are only
interested in the leading terms of the free energy, which we compare
with those of the infinite quasiperiodic tiling. We find that, at
least for the leading orders in $w$, the rational coefficients of the
approximants converge rapidly towards the irrational coefficients
obtained for the quasiperiodic tiling.

In conclusion, it is doubtful whether the computational effort
necessary to extend the expansions to higher order will result in a
considerable improvement of the estimates of the critical properties.
Instead, it might be more rewarding to consider periodic approximants
and use methods as those outlined in \sref{sec6} and \cite{Vdo2}
to compute physical quantities, such as for instance the magnetization
or correlation functions.

\ack

We thank M.~Baake for discussions and useful comments.  Financial
support by Deutsche Forschungsgemeinschaft (DFG) is gratefully
acknowledged.

\end{document}